\documentclass[twocolumn,showpacs,aps,amsmath,amssymb]{revtex4}
\usepackage[dvips]{graphicx}

\newcommand{\Prob}{{\mathrm{Pr}}}

\newcommand{\I}{{\mathrm{i}}}

\newcommand{\ket}[1]{|#1\rangle}
\newcommand{\bra}[1]{\langle#1|}
\newcommand{\proj}[1]{\ket{#1}\bra{#1}}

\newcommand{\smfrac}[2]{\mbox{$\frac{#1}{#2}$}}

\newcommand{\Tr}{\operatorname{Tr}}

\begin{document}
\title{Optimal phase estimation in quantum networks}
\author{Wim van Dam}
\affiliation{Departments of Computer Science and Physics,
University of California, Santa Barbara, Santa Barbara, CA
93106-5110 United States of America}
\author{G. Mauro D'Ariano}
\affiliation{Dipartimento di Fisica ``A. Volta'' and CNISM, via Bassi 6, 27100 Pavia, Italy}
\author{Artur Ekert}
\affiliation{Mathematical Institute, University of Oxford,
24-29 St Giles', Oxford, OX1 3LB, UK and Department of Physics,
National University of Singapore, 2 Science Drive 3, Singapore
117542}
\author{Chiara Macchiavello}
\affiliation{Dipartimento di Fisica ``A. Volta'' and CNISM, via Bassi 6, 27100 Pavia, Italy}
\author{Michele Mosca}
\affiliation{Institute for Quantum Computing, University of
Waterloo, N2L 3G1, St. Jerome's University, N2L 3G3, and Perimeter
Institute for Theoretical Physics, N2L 2Y5, Waterloo, ON, Canada}
\begin{abstract}
We address the problem of estimating the phase $\phi$ given $N$
copies of the phase rotation $u_{\phi}$ within an array of quantum
operations in finite dimensions. We first consider the special case
where the array consists of an arbitrary input state followed by any
arrangement of the $N$ phase rotations, and ending with a POVM. We
optimise the POVM for a given input state and fixed arrangement.
Then we also optimise the input state for some specific cost
functions. In all cases, the optimal POVM is equivalent to a quantum
Fourier transform in an appropriate basis. Examples and applications
are given.
\end{abstract}
 \maketitle

\section{Introduction}

Extracting information that is encoded in the relative phase of
quantum systems is exploited in several quantum information
processing tasks and applications. For example, most existing
quantum algorithms with super-polynomial speed-up involve phase
estimation \cite{CEMM} and more recently it was shown that one
natural formulation of the phase estimation problem is BQP-complete
\cite{wzBQP}.  Moreover, information is encoded into phase
properties in some quantum cryptographic protocols \cite{BB84}, and
in schemes on which atomic clocks are based \cite{atclocks}.
Therefore, the issue of estimating the phase in the most efficient
way is of fundamental importance.

In this paper we address the problem of optimally estimating the
phase $\phi$ in a quantum network of qubits given $N$ copies of the
phase rotation $u_{\phi}=\exp(i\proj{1}\phi)$, where $\ket{0}$ and
$\ket{1}$ denote the computational basis of a single qubit.

There is a variety of relevant scenarios that can be considered. For
example, the most general scenario, where one can prepare any
initial state, and apply an arbitrary quantum circuit including the
$N$ phase rotation gates, is considered in \cite{DDEMMa} for
arbitrary cost functions, and in \cite{GLM06} for a specific cost
function.  In \cite{DDEMMa} it is shown that for any cost function,
the general case can be reduced to optimizing the initial state of a
procedure with a very specific form. More specific scenarios are
relevant, for example, when performing state estimation (e.g.\
\cite{dbe}) or phase estimations in other special circumstances,
such as in the final step of a dihedral hidden subgroup algorithm
\cite{eh}.

The paper is organised as follows. In Sect.~\ref{s:problem} we
introduce these specific phase estimation scenarios, where the phase
shift gates are applied in a special way. In Sect. \ref{s:ope} we
optimise the phase estimation procedure in these cases, by first
deriving the optimal POVM for a fixed initial state and cost
function, and then by optimising the average cost also on the form
of the initial state. In Sects.~\ref{s:N}, \ref{s:shor} and
\ref{s:HSP} we give some examples of interest where our results can
be applied, namely the case of phase estimation in a system of $N$
identically prepared qubits, the Shor algorithm and the dihedral
hidden subgroup problem. Finally, in Sect.~\ref{s:conc} we summarise
the main results and discuss their implications. The paper ends with
some appendices where some details of the derivations presented in
the text are reported.

\section{Special case}
\label{s:problem}

In this section, we will study a special case of the general problem
of optimal phase estimation, where we restrict attention to one
particular way of applying the phase rotation operator $u_{\phi}$
some finite number of times, and we also restrict attention to a
special (but widely used) class of ``cost functions''.

In particular, we consider the task of estimating the unknown phase
shift parameter $\phi\in[0,2\pi]$ of a unitary transformation
$U_\phi$ acting on $L$ qubits (with Hilbert space ${\cal H}^{\otimes
L}$, ${\cal H}\simeq C^2$), with $U_\phi$  of the form
\begin{equation}
U_\phi=\otimes_{l=1}^L u_\phi^{n_l} \;, \label{uphi}
\end{equation}
where $u_\phi$ is the elementary single qubit phase-shift gate
defined in the introduction.
More explicitly the $l$th qubit undergoes the unitary phase shift
\begin{align}
\ket{0}_l  & \mapsto \ket{0}_l,\\
\ket{1}_l & \mapsto e^{in_l\phi}\ket{1}_l,
\end{align}
where $\{\ket{0}_l,\ket{1}_l\}$ is a basis for the $l$-th qubit and
$n_l$ is an integer number, with the constraint $\sum_l n_l=N$.
\par
\begin{figure}[htb]
\begin{center}
\includegraphics[width=8cm]{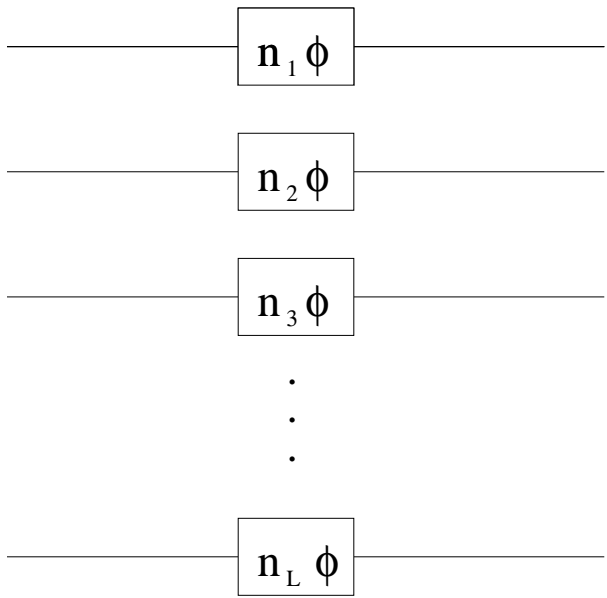}
\end{center}
\caption{Network representation of the action of the unitary
operator acting on $L$ qubits $U_\phi=\otimes_{l=1}^L u_\phi^{n_l}$, with
$n_l$ integers. Each box represents a single qubit unitary phase
shift operation.} \label{f:net}
\end{figure}

 The operator $U_\phi$, $2\pi$-periodic with respect to $\phi$, 
operates locally on each of the $L$ qubits,
and in the Schr\"{o}dinger picture it acts on a known initial state
$\ket{\Psi_0}\in{\cal H}^{\otimes L}$. The general scenario is
illustrated in Fig.~\ref{f:net}.
\par The problem we want to address can be phrased as follows:
\begin{itemize}
\item[] {\em Given the set of numbers $\{n_l,l=1,\dots, L\}$,
with $\sum_l n_l=N$, we want to find the best measurement
    procedure to estimate the phase $\phi$. }
\end{itemize}
\par This estimation problem is posed in the general framework of {\em quantum estimation theory}
\cite{helstrom}. One first defines a {\em cost function}
$C(\phi_*,\phi)$, which assesses ``the cost one has to pay'' for
errors in the estimated value $\phi_*$ when the true one is $\phi$
(such a function is typically a non decreasing function of
$|\phi-\phi_*|$ so that we pay more for larger errors in the
estimate). For any given cost function $C(\phi_*,\phi)$ and \emph{a
priori} probability distribution $p_0(\phi)$ for the true value
$\phi$ one then evaluates the average cost in the estimate
\begin{equation}
\bar C = 
\int_0^{2\pi}d\phi\,p_0(\phi)\int_0^{2\pi}d\phi_*C(\phi_*,\phi) \,p(\phi_*|\phi)\;,
\label{avc}
\end{equation}
where $p(\phi_*|\phi)$ is the conditional probability
 of estimating $\phi_*$ given the true value $\phi$.
The conditional probability is given by the Born rule
\begin{equation}
p(\phi_*|\phi) = 
\Tr[D_{\phi_*} U_\phi\ket{\Psi_0}\bra{\Psi_0}
U_\phi^\dagger]\;, \label{prob}
\end{equation}
where $D_{\phi_*}$ denotes the POVM density \cite{helstrom} of the
estimation, with $D_{\phi_*}$ a positive operator, normalized as
\begin{equation}
\int_0^{2\pi}d\phi_* D_{\phi_*}=I,
\end{equation}
$I$ denoting the identity operator on ${\cal H}^{\otimes L}$. The
estimation problem consists in finding the optimal POVM which
minimizes the average cost $\bar C$ in Eq.~(\ref{avc}).
\par We assume that $\phi$ is \emph{a priori} uniformly distributed in $[0,2\pi]$
with probability density $p_0(\phi)=\frac{1}{2\pi}$ (see
\ref{app_uniform_prior} for an explanation). Moreover, we want to
have no privileged phase value in the estimation, hence the error is
paid only as a function of the distance between the true value and
the estimated one, so that the cost function is actually an even
function of a single variable $C(\phi_*,\phi)\equiv C(\phi_*-\phi)$.
In the following we will consider some examples of cost functions
corresponding to relevant optimization criteria.

For a cost function that depends only on $\phi_*-\phi$, one can
prove \cite{holevo} that there exists an optimal POVM which is
covariant, namely
\begin{equation}
d\phi_* D_{\phi_*}=U_{\phi_*} \xi U_{\phi_*}^\dagger
\frac{d\phi_*}{2\pi}\;, \label{dmu}
\end{equation}
where $\xi\equiv 2\pi D_0$ is a positive operator, corresponding to
a conditional probability $p(\phi_*|\phi)$ that will also depend
only on $\phi_*-\phi$. Therefore, using Eq.~(\ref{prob})
and the invariance of the trace under cyclic permutations one has
\begin{align}
p(\phi_*|\phi) & \equiv p(\phi_*-\phi)\\
& =\Tr[U_\phi^\dagger
D_{\phi_*}U_\phi \ket{\Psi_0}\bra{\Psi_0}]\\
 & =\Tr[U_{\phi-\phi_*}^\dagger D_0 U_{\phi-\phi_*}
\ket{\Psi_0}\bra{\Psi_0}] . 
\end{align}
In other words, the POVM is generated from a positive operator $\xi$ under the
action of the unitary Abelian group of the operators $U_{\phi_*}$,
with $\phi_*\in[0,2\pi]$. Hence the problem is reduced to finding
the positive operator $\xi$ that minimises the cost $\bar C$ in Eq.
(\ref{avc}).

It is convenient to seek the solution of the optimization problem
using the representation of the {\em shift operator} $K$ defined
as
\begin{equation}
U_\phi=\exp(i K\phi).
\end{equation}
The operator $K$ has integer (possibly degenerate) spectrum that is
contained in ${\mathbb Z}_q$ for $q=N+1$ (the $2\pi$-periodicity of
$U_\phi$ in Eq.~(\ref{uphi}) implies integer values for the spectrum
of $K$). In the following we will denote by $P_k$ the projector over
the eigenspace of $K$ corresponding to the eigenvalue $k\in{\mathbb
Z}_q$. We introduce a (generally incomplete) set of orthonormal
states $\{\ket{k}\}$ uniquely defined as follows. For $k \in
{\mbox{spec}} K$, define $p_k = \bra{\Psi_0} P_k \ket{\Psi_0}$ and
for those $k$ such that $p_k \neq 0$ let
\begin{equation}
\ket{k} = \frac{P_k\ket{\Psi_0}}{\sqrt{p_k}}
\end{equation}
(recall that the initial state of the $L$ qubits $\ket{\Psi_0}$ is
known by hypothesis). In the following we will denote by ${\mathbb
S}\subseteq {\mathbb Z}_q$ the set of eigenvalues $k$ such that
$P_k\ket{\Psi_0}\neq 0$.
\par The problem of state estimation is
now restricted to the Hilbert space ${\cal
K}=\operatorname{Span}\{\ket{k},\, k\in {\mathbb S}\}$, with
$\mbox{dim}({\cal K})\le q$. Therefore, without loss of
generality, the POVM density can be completed in the block
diagonal form
\begin{equation}
D_\phi=D^{\cal K}_\phi\oplus D^{{\cal K}^\perp}_\phi
\end{equation}
on ${\cal H}^{\otimes N}={\cal K}\oplus{\cal K}^\perp$,  with
$D^{{\cal K}^\perp}_\phi$ any arbitrary POVM density on ${\cal
K}^\perp$ (the component of the POVM density $D^{{\cal
K}^\perp}_\phi$  acting on the subspace ${{\cal K}^\perp}$ can be
arbitrary because the state $\ket{\Psi_0}$ has vanishing
projection on this subspace and therefore this component does not
give any contribution to the average cost).
\par The above procedure has been designed to deal with the (possible)
degeneracy
of the operator $K$, reducing the problem to the ``canonical'' phase
estimation problem for non degenerate shift operator $H$ on the
space ${\cal K}$
\begin{equation}
\ket{\Psi_0} \mapsto\exp(iH\phi)\ket{\Psi_0}, \quad 
H =\sum_{k\in{\mathbb
S}} k \ket{k}\bra{k} \label{restr}
\end{equation}
with initial state
\begin{equation}
\ket{\Psi_0}=\sum_{k\in{\mathbb S}} x_k\ket{k}.
\end{equation}
Notice that due to arbitrariness of the POVM density complement
$D_\phi^{{\cal K}^\perp}$ on ${\cal K}^\perp$, it is always
possible to complete ${\mathbb S}$ to ${\mathbb Z}_q$, without
changing the optimality of the POVM.

Note that this implies that the optimal POVM depends only on the set
${\mathbb S}$, which is contained in the set of integers which can
be expressed as $\sum_i b_i n_i$, for some $b_i \in \{0,1\}$ (also
referred to as ``subset sums'' of the set $\{n_1, n_2,\dots,
n_L\}$). The choice of the initial state $\ket{\Psi_0}$ determines
the weight $x_k$ of each possible integer $k$ of this form, and we
define ${\mathbb S}$ to be the $k$ with $x_k \neq 0$.

This means that if we were given the freedom to choose the initial
state $\ket{\Psi_0}$, then all the partitions of $N$ into parts
$n_1, n_2,\dots, n_L$, with the same subset sums 
lead to the same optimal expected cost for estimating $\phi$.

Given full freedom to partition $N$, one can always achieve all
possible subset sums between $0$ and $N$. One obvious way to achieve
this is to have $L=N$ and $n_1 = n_2 = \dots = n_N = 1$. However
this requires $N$ qubits. One call also achieve all subset sums with
$O(\log_2 N)$ qubits by letting $n_i = 2^i$, for $0 \leq i < L =
\lfloor \log_2 N \rfloor $, and $n_L = N - 2^{L}+1$.  A simple basis
change allows one to assume that $\ket{k}$ is a tensor product state
representing the integer $k$ expressed in binary (described in
 \ref{app_exponentiate} and illustrated in Fig.~\ref{f:exponentiate}).

\section{Optimal phase estimation in the special case}

\label{s:ope}

In this section we will review the derivation of the
optimal procedure to estimate the phase $\phi$ in Eq.~(\ref{restr})
\cite{holevo}. We will first derive in subsec.~\ref{s:optpom} the
optimal POVM for a fixed initial state $\ket{\Psi_0}$ and for a
given cost function. As mentioned above, the optimal POVM is the one
that minimises the average cost in Eq.~(\ref{avc}). We will show
that a general solution can be found, that depends on the form of
the initial state $\ket{\Psi_0}$ only through the orthonormal set
$\{\ket{k},\, k\in{\mathbb S}\}$ and that holds for a large class of
cost functions which includes for example the fidelity and the
window function as special cases. Later, in subsec.~\ref{s:optstates}, 
we will further optimise the solution by deriving
the form of the initial state $\ket{\Psi_0}$ that leads to the
minimum average cost for a chosen cost function and using the
optimal POVM. In this way we give a prescription to prepare the
initial state of the $L$ qubits in order to have the most efficient
phase estimation for a given network (i.e.\ for a given set of
integer numbers $\{n_l,l=1,\dots, L\}$) and for a given cost
function.

\subsection{Optimal POVM}
\label{s:optpom}
In order to find the optimal phase estimation procedure we have to
derive the positive operator $\chi=2\pi D_0^{\cal K}$ on the
subspace ${\cal K}$, that minimizes the cost $\bar C$ in Eq.
(\ref{avc}). We will conveniently use the representation of $\chi$
in the $\ket{k}$ basis, namely
\begin{equation}
\chi=\sum_{h,k\in{\mathbb S}}\ket{h}\langle k| \chi_{hk}\;.
\label{xi}
\end{equation}
As a cost function we consider a generic $2\pi$-periodic even
function, which can be expressed as a Fourier series in the form
\begin{equation}
C(\phi)=-\sum_{l=0}^{\infty}c_l\cos l\phi.
\end{equation}
The average cost can then be written as
\begin{align}
\bar C=&-c_0 -\sum_{l=1}^\infty c_l\int \frac{d\phi}{2\pi} \cos
l\phi\\& \times \sum_{hk\in{\mathbb S}}
e^{i(h-k)\phi}\chi_{hk}\langle k\ket{\Psi_0}
\langle{\Psi_0}\ket{h}\;. \label{avcxi1}
\end{align}
By calculating the integrals the above expression can be reduced
to
\begin{equation}
\bar C = -c_0-\frac{1}{2}\sum_{l=1}^{q-1} c_l \sum_{h,k\in{\mathbb
S},\;|h-k|=l} \langle{\Psi_0}\ket{h}\chi_{hk}\langle
k\ket{\Psi_0}\;. \label{avcxi2}
\end{equation}
Notice that the sum over $l$ is truncated by the vanishing of the
$\ket{\Psi_0}$ components. Positivity of the operator $\chi$ implies
that \cite{holevo}
\begin{equation}
|\chi_{hk}|\leq\sqrt{\chi_{hh}\chi_{kk}}=1\;, \label{mat}
\end{equation}
where the last equality $\chi_{hh}=1$ $\forall h$ is a consequence
of the POVM completeness condition $\int d\phi D_\phi^{\cal K}=
I_{{\cal K}}$ ($I_{{\cal K}}$ is the identity operator on the
Hilbert space ${\cal K}$). By exploiting Eq.~(\ref{mat}) we can
now write the following inequality
\begin{equation}
\begin{split}
{\mbox{sign}}(c_l) &\sum_{h,k\in{\mathbb
S},\;|h-k|=l}\langle{\Psi_0}\ket{h}\chi_{hk}\langle k\ket{\Psi_0}
\\ &\leq
\sum_{h,k\in{\mathbb S},\;|h-k|=l}|\langle{\Psi_0}\ket{h}||\langle
k\ket{\Psi_0}|\;. \label{inpom}
\end{split}
\end{equation}
The equality in the above equation is achieved only for
$\chi_{hk}={\mbox{sign}}(c_{|h-k|})$ (we recall that
$\langle\Psi_0\ket{k}> 0$ $\forall k\in {\mathbb S}$ by
construction). The average cost is minimised when the equality in
Eq.~(\ref{inpom}) is satisfied and the minimum value is given by
\begin{equation}
\bar C=-c_0 -\frac{1}{2}\sum_{l=1}^{q-1} |c_l|
\sum_{h,k\in{\mathbb S},\;|h-k|=l}|\langle{\Psi_0}\ket{h}||\langle
k\ket{\Psi_0}| \label{minc}
\end{equation}
where we can set ${\mbox{sign}}(0)=1$, since the cost $\bar C$ is
independent of $\chi_{hk}$ for $c_{|h-k|}=0$. Notice, however, that
positivity of the operator $\chi$ in the form
$\chi_{hk}={\mbox{sign}}(c_{|h-k|})$ is not guaranteed for an
arbitrary choice of the coefficients $c_l$. For the rest of Section
\ref{s:ope} we will restrict to the particular form of coefficients
$c_l\geq 0$ $\forall l\geq 1$ considered by Holevo \cite{holevo}. In
this case $\chi$ has all unit elements $\chi_{hk}=1$ and is
positive. The optimal POVM corresponding to Eq.~(\ref{dmu}) can be
written as follows
\begin{equation}
d\phi\, D_\phi^{\cal K}\equiv\frac{d\phi}{2\pi}
\ket{e^{i\phi}}_{\cal K} {}_{\cal K}\bra{e^{i\phi}} \label{pomopt}
\end{equation}
where
\begin{equation}
\ket{e^{i\phi}}_{\cal K}= \sum_{k\in{\mathbb S}}
e^{ik\phi}\ket{k}.
\end{equation}

We want to stress that the Holevo condition on the Fourier
coefficients is not very restrictive. In fact, it corresponds to a
large class of optimization criteria. It includes for example the
likelihood criterion for $C_L(\phi)=-\delta_{2\pi}(\phi)$, the
periodicized variance for $C_V(\phi)=4\sin ^2 (\phi/2)$ and the
fidelity optimization $F(\phi)=|\bra{\Psi_0}U_\phi\ket{\Psi_0}|^2$
with $C_F(\phi)=1-F(\phi)$ and $c_l=2\sum_{h,k\in{\mathbb
S},\;|h-k|=l}x_h^2 x_k^2$ where $x_k=\bra{k}\Psi_0\rangle$. 
Notice that also a cost function often considered implicitly in
computer science, namely the ``window function'' $C_W(\phi)=0$ for
$|\phi|\leq\epsilon$, $C_W(\phi)=1$ for $|\phi|>\epsilon$, defined for
$\phi\in[-\pi,+\pi]$, can be included in this class under some
restrictions. Actually, the Fourier coefficients for such a function
take the form $c_0=\frac{\epsilon}{\pi}-1$ and
$c_l=\sin(l\epsilon)/l\pi$ for $l>0$, and therefore they are all
positive for $0<l< q$ when the width of the window function satisfies
the condition $\epsilon\leq \pi/q$. Since the Fourier coefficients
that contribute to the average cost in Eq.  (\ref{avcxi2}) correspond
to $l< q$ and are all positive (apart from $c_0$ but this does not
affect the optimisation of the POVM), the optimal POVM for the window
function with $\epsilon\leq \pi/q$ is still given by
Eq.~(\ref{pomopt}). We want to point out that the same POVM
(\ref{pomopt}) would optimize the average cost for Holevo type cost
functions even for a certain class of mixed states \cite{nota}.

The optimal POVM in the form (\ref{pomopt}) does not correspond to a
straightforward physical measurement scheme (that is, one with
finite resources, including finite resolution) because it gives a
continuous value of the phase as the estimated value. We will now
find a more convenient discrete description.
\par  We can first complete
the set ${\mathbb S}$ to ${\mathbb Z}_q$ by appropriate choice of
the arbitrary density $D_\phi^{{\cal K}^\perp}$, extending ${\cal
K}$ to the span of those vectors $\ket{k}$ corresponding to
$P_k\ket\Psi_0=0$, and restricting ${\cal K}^\perp$ accordingly,
without changing the optimality of the POVM. The optimal relevant
POVM density is then given by Eq.~(\ref{pomopt}), with
\begin{equation}
\ket{e^{i\phi}}_{\cal K}= \sum_{k=0}^{q-1} e^{ik\phi}\ket{k}.
\end{equation}
Due to the covariance of the optimal POVM and the uniform prior
distribution $p_0(\phi)$, the average cost $\bar C$        in Eq.
(\ref{avcxi1}) takes exactly the same value if the continuous POVM
(\ref{pomopt}) is restricted to only a set of $q$ equally spaced
values $\phi_s=\frac{2\pi}{q}s$, $s=0,1,\dots,q-1$. This can be easily
proved using  the identity
\begin{equation}
\delta_{n0}=\int_0^{2\pi}\frac{d\phi}{2\pi}e^{in\phi}=
\frac{1}{q}\sum_{s=0}^{q-1}e^{in\phi_s}
\end{equation}
with $n\in {\mathbb Z}_q$. Therefore, for every covariant POVM there exists
always a discrete POVM giving the same average cost. In the present case of
Holevo cost the optimal POVM becomes the orthogonal projector-valued measure
\begin{equation}
E_s=\ket{\phi_s}\langle{\phi_s}| \label{pomort}
\end{equation}
where
\begin{equation}
\ket{\phi_s}=\frac{1}{\sqrt{q}}\sum_{k=0}^{q-1}
e^{ik\phi_s}\ket{k}\; \label{phis}
\end{equation}
are orthogonal states for $s=0,\dots,q-1$. Notice that from the form
of states (\ref{phis}) one can see that this estimation procedure
can be implemented by the customary discrete quantum Fourier
transform (QFT) network. Exact implementation of the QFT for any
positive integer $q$ was detailed in \cite{mz}.

\subsection{Optimal states}
\label{s:optstates}

We will now perform a further optimisation, namely we derive the
form of the initial state $\ket{\Psi_0}$ that leads to the minimum
average cost (how to generate such initial states is described in
\cite{km}). As we will see soon, this step depends crucially on the
form of the cost function, which has to be specified in advance. We
first notice that for a covariant POVM the average cost can be
recast in the following form
\begin{equation}
\bar C=\Tr[\hat C\proj{\Psi_0}] \label{co}
\end{equation}
where the cost operator is defined as
\begin{equation}
\hat C=\int_0^{2\pi} d\phi\, D_\phi^{\cal K} C(\phi)\;.
\label{co2}
\end{equation}
Without loss of generality we will take ${\mathbb S}\equiv{\mathbb
Z}_q$ in the following. (Notice that one can define the operator
$\hat C$ with the complete POVM density $D_\phi$ over ${\cal
H}={\cal K}\oplus{\cal K}^\perp$, since the term $D_\phi^{{\cal
K}^\perp}$ will not contribute to the trace (\ref{co})).

The average cost $\bar C$ can be minimized over the coefficients
of $\ket{\Psi_0}$ by using the Lagrange multipliers method to
account for the normalization constraint. One has to minimize the
bilinear function
\begin{equation}
{\cal L}[\Psi_0]=\bra{\Psi_0}\hat
C\ket{\Psi_0}-\lambda\bra{\Psi_0} \Psi_0\rangle \label{elle}
\end{equation}
which gives the eigenvalue equation
\begin{equation}
\hat C\ket{\Psi_0}=\lambda|\Psi_0\rangle
\end{equation}
where now the Lagrange parameter $\lambda$ plays the role of an
eigenvalue. As mentioned above, we have now to specify the form of
the cost function. We will consider two particular cases which
correspond to the optimal POVM (\ref{pomopt}). We first consider the
cost function $C_V(\phi)=4\sin^2(\phi/2)$ (more details are given in
 \ref{ap:maxfid}  ). In this case the cost operator in 
Eq.~(\ref{co2}) takes the form
\begin{equation}
\hat C_V=e_+ + e_- -2
\end{equation}
where
\begin{equation}
e_+=\sum_{k=0}^{q-2}\ket{k+1}\bra{k},\quad e_-=e_+^\dagger.
\end{equation}
In terms of the coefficients $x_k=\bra{\Psi_0}k\rangle >0$ the
eigenvalue equation becomes the recurrence relation
\begin{equation}
x_{k+1}(\lambda)+x_{k-1}(\lambda)-(2+\lambda)x_k(\lambda)=0
\end{equation}
with boundary conditions $x_{-1}(\lambda)=x_q(\lambda)=0$. The
solution can be written in terms of the Chebyshev polynomials of
the second kind \cite{TABLE}, leading to
\begin{equation}
x_{j}=\sqrt{\frac{2}{q+1}}\sin\left(\frac{j+1}{q+1}\pi\right)\;.
\label{optst}
\end{equation}
The minimum cost in this case is given by
\begin{equation}
\bar C_V=\sin ^2\left(\frac{\pi}{2(q+1)}\right)\;.
\label{mincv}
\end{equation}

As a second example we consider the window function cost
$C_W(\phi)$ defined in the previous subsection with the condition
$\epsilon\leq \pi/q$, so that the optimal POVM is still given by
Eq.~(\ref{pomopt}). In this case the cost operator takes the form
\begin{equation}
\hat C_W=\left(1-\frac{\epsilon}{\pi}\right)-\frac{1}{2\pi}\sum_{l=1}^{q-1}
\frac{\sin l\epsilon}{l}(e_+^l+e_-^l)\;,
\end{equation}
where
\begin{equation}
e_+^l=\sum_{k=0}^{q-l-1}\ket{k+l}\bra{k},\quad
e_-^l={(e_+^l)}^\dagger.
\end{equation}
The eigenvalue equation has the form
\begin{equation}
x_{k}\left(\lambda+\frac{\epsilon}{\pi}-\right)+
\frac{1}{2\pi}\sum_{m(\neq
k)=0}^{q-1}\frac{\sin(k-m)\epsilon}{k-m}x_m =0\;.
\end{equation}
This has an easy solution for very narrow window functions, such
that $\epsilon\ll 1/q$. In this case the coefficients
$\sin(m\epsilon)/m$ can be approximated by a constant independent
of $m$ and the solution for the optimal state $\ket{\Psi_0}$ is
simply given by the equally weighted state, i.e.\
\begin{equation}
x_k=1/\sqrt{q}, \quad k=0,\dots,q-1\;. \label{optstW}
\end{equation}
The minimum cost in this case takes the form
\begin{equation}
\bar C_W=\left[1-\frac{\epsilon}{2\pi}(q+1)\right]\;.
\end{equation}

\section{Example 1: phase estimation of $N$ identically prepared qubits}
\label{s:N}

We specify here the phase estimation problem to a particular case:
estimation of the phase shift undergone by $N$ qubits initially in
the same state, namely
\begin{equation}
U_\phi (a\ket{0}+b \ket{1})^{\otimes N}= (a\ket{0}+b
e^{i\phi}\ket{1})^{\otimes N}, \label{ncopies}
\end{equation}
i.e.\ the unitary operator $U_\phi$ is given by
\begin{equation}
U_\phi
=\otimes_{l=1}^N\exp\left[\frac{i}{2}(\sigma_l^z-1)\phi\right]\;,
\label{uphincopies}
\end{equation}
where $\sigma_l^z$ represents the Pauli operator for the $l$-th
qubit. In the network language of Fig.~\ref{f:net} this case
corresponds to $n_l=1\;\;\forall l$. One can see that for both
$a\neq 0$ and $b\neq 0$ the $\ket{k}$ basis introduced in 
subsec.~\ref{s:optpom} is given by
\begin{align}
\ket{k}&=e^{i\phi_k}\ket{k}_{\mbox{sym}}\\ 
& \equiv e^{i\phi_k}
\left(\begin{array}{c}N\\k\end{array}\right)^{-1/2}
\sum_{\{s_i=0,1\}} \delta(\sum_i s_i-k)\otimes_{i=1}^N\ket{s_i},
\end{align}
namely it is the basis of the symmetric multiplet with the choice of
phases $\phi_k=(N-k){\mbox{arg}}(a)+k\, {\mbox{arg}}(b)$ (in this
case $q=N+1$ because the subspace ${\cal K}$ is the symmetric
subspace of the $N$ qubits). The optimal POVM restricted to the
symmetrical tensor subspace ${\cal K}=\left({\cal H}^{\otimes
N}\right)_+$.

is given in Eq.\~(\ref{pomort}) for cost functions of the Holevo
class.

The optimized state among the ones of the form (\ref{ncopies}) for the
cost function $C_V(\phi)=4\sin^2\phi/2$, and also for the fidelity
error $1-F(\phi)$ with $F(\phi)=|a|^4+|b|^4+2|a|^2|b|^2\cos(\phi)$, is
given by $a=b$ and leads to the minimum cost
\begin{equation}
C_V=\frac{1}{2}-\frac{1}{2^{N+1}}\sum_{k=0}^{N-1}
\sqrt{\left(\begin{array}{c}N\\k\end{array}\right)
\left(\begin{array}{c}N\\k+1\end{array}\right)}\;.
\end{equation}
This result has also been reported in Ref.~\cite{dbe}.

Notice that we can also phrase the problem in a wider sense and
ask for the optimal strategy to estimate the phase shift generated
by the unitary operator (\ref{uphincopies}) on a generic symmetric
state of the $N$ qubits and optimise the average cost with respect
to the initial state, as in subsection~\ref{s:optstates}. It is
interesting to notice that in this case the optimal states for the
cost function $C_V$ and the window cost function, given in 
Eqs.~(\ref{optst}) and (\ref{optstW}) respectively, are entangled
states of the $N$ qubits.

Notice also that for $U_\phi$ of the form (\ref{uphincopies}) we can
solve an even more general problem, where the initial state
$\ket{\Psi_0}$ does not belong to the symmetric subspace. In this
case, in fact, the best POVM is of the form $Z^\dagger E_s Z$, where
$\ket{\Psi_0}=Z\ket{\Lambda_0}$ and $\ket{\Lambda_0}\in\left( {\cal
H}^{\otimes N}\right)_+$, the symmetric subspace of the $N$ qubits
(one can always find a suitable $Z$ commuting with a $U_\phi$ of the
form in Eq.~(\ref{uphincopies})).

\section{Example 2: the Shor algorithm}
\label{s:shor}

In this section we will consider another particular case,
corresponding to the phase estimation as the final step of the Shor
algorithm in the formulation given in Ref.~\cite{CEMM}. This step
was also optimized in \cite{bessen}. The phase operator $U_\phi$ in
this case has the form
\begin{equation}
U_\phi =\otimes_{l=1}^L u_\phi^{2^{l-1}}.
\end{equation}
This corresponds to the network representation of Fig.~\ref{f:net}
with $n_l=2^{l-1}$. In this case the problem does not have
degeneracy and the subspace ${\cal K}$ is the whole Hilbert space
${\cal H}^{\otimes L}$ of the $L$ qubits, namely $q=2^L$. The
eigenvectors of the operator $K$ are simply given by the
computational basis of the $L$ qubits
\begin{equation}
\ket{k}=\otimes_{l=1}^L\ket{s_l}\,,\quad k=\sum_{l=1}^L s_l
2^{l-1}\;.
\end{equation}
For cost functions of the Holevo class the optimal POVM in Eq.
(\ref{pomort}) here corresponds to the Quantum Fourier Transform
measurement discussed in Ref.~\cite{CEMM}, where a network
realisation is also given. Here, our general method proves that
this measurement procedure is optimal.

Regarding the optimized states in Eq.~(\ref{optst}) for the cost
function $C_V$ one should notice that the state is partially
entangled. To be useful in practice, we would also need to consider
the computational complexity of creating this state for use as a
possible initial state for the Shor algorithm.

If the window cost function is considered, in the limit of very
small width analysed in the previous section, the optimal state is
given by
\begin{equation}
\ket{\Psi_0}= \left[\tfrac{1}{\sqrt
2}(\ket{0}+\ket{1})\right]^{\otimes L},
\end{equation}
namely it is a factorised state of the $L$ qubits, as the one
considered in the scheme of Ref.~\cite{CEMM}.

\section{Example 3: Dihedral Hidden Subgroup Problem}
\label{s:HSP}

The hidden subgroup problem (HSP) is the problem of finding
generators for a subgroup $K$ of a group $G$ given a black-box that
implements a function $f:G \rightarrow X$ satisfying $f(x) = f(y)
\Leftrightarrow x - y \in K$. In other words, $f$ is constant on
cosets of $K$ and distinct on different cosets.

Assuming a reasonable presentation of the group $G$, there is an
efficient quantum algorithm for solving the HSP in Abelian groups.
This algorithm is a natural generalization of Shor's algorithm
\cite{shor} and can be cast as a phase estimation problem as
outlined in \cite{CEMM} (based on the approach of \cite{kitaev}).
There has been limited success in solving the HSP for non-Abelian
groups. See \cite{miklos} for recent results and references.
Ettinger and H{\o}yer \cite{eh} reduced the HSP for the dihedral
group to the following phase estimation problem:

Given a polynomial (in $n$) number of qubits of the form $\ket{0} +
e^{i k_j \phi} \ket{1}$ and the integers $k_j$, where the $k_j$ are
selected uniformly at random from $\{0,1,2,\dots, 2^n - 1\}$,
estimate $\phi$ with error at most $\frac{1}{2^n}$.

Ettinger and H{\o}yer showed that an optimal measurement would
solve the HSP with high probability.

Our results from Section \ref{s:problem} tell us the optimal POVM.
Let $S= \sum_{j=1}^n k_j$. Let $S_j$ equal the set of solutions
$(b_1, b_2,\dots, b_n) \in \{0,1\}^n$ to the equation $\sum_{i}
b_i k_i = j$, and let $n_j = |S_j|$ equal the number of such
solutions. For $j$ with $n_j \neq 0$, let $\ket{S_j} =
\frac{1}{\sqrt{n_j}} \sum_{ b_1 b_2 \dots b_n | \sum_{i} b_i k_i =
j} \ket{ b_1 b_2 \dots b_n}$.

Note that 
\begin{equation} 
(\ket{0} + e^{i k_1 \phi}\ket{1})\otimes \cdots \otimes
(\ket{0} + e^{i k_n \phi} \ket{1})
= \sum_{j = 0}^{S} \frac{\sqrt{n_j}}{2^n} \ket{S_j} .
\end{equation}

Thus the optimal POVM could be achieved by first performing a
unitary basis change $U$ that maps $\ket{S_k} \mapsto \ket{k}$,
where the integer $k$ is represented as a binary string in the
computational basis, followed by a quantum Fourier transform in the
computational basis, and a measurement in the computational basis.

However, implementing $U$ would solve the subset sum problem, which
is known to be NP-complete (since $U^{-1}$ maps $\ket{k}$ to a
uniform superposition of strings representing subsets with sum $k$).
However, such a measurement is optimal, and a sufficiently good
approximation of this measurement would suffice.
Regev \cite{regev} showed that it suffices to be able to solve the
subset sum problem on average in order to find a sufficiently
precise estimate of $\phi$ that allows one to solve the Hidden
Subgroup Problem for the Dihedral group.

\section{Discussion}
\label{s:conc}

In this paper we have addressed the problem of finding the optimal
estimating procedure for the real parameter $\phi$ given $N$ copies
of the single qubit phase rotation $u_{\phi}$ within a quantum
network in finite dimensions. We have derived the optimal
measurement procedure in the special case where the network consists
of an arbitrary input state followed by any arrangement of the $N$
phase rotations, and found also the optimal states corresponding to
some cost functions of interest.

This result is general and can be applied to many cases of interest.
In particular, we have considered the phase estimation problem as
the final step of the Shor algorithm in the formulation given in
Ref.~\cite{CEMM} and we have proved that the quantum Fourier
transform performed in that case is indeed the optimal phase
estimation procedure. As another example, we have also shown how our
result can be applied to the dihedral hidden subgroup problem.

\appendix

\section{Phase shift circuit} \label{app_exponentiate}

\begin{figure*}[htb]
\begin{center}
\includegraphics[scale=0.75,bb=150 85 450 450]{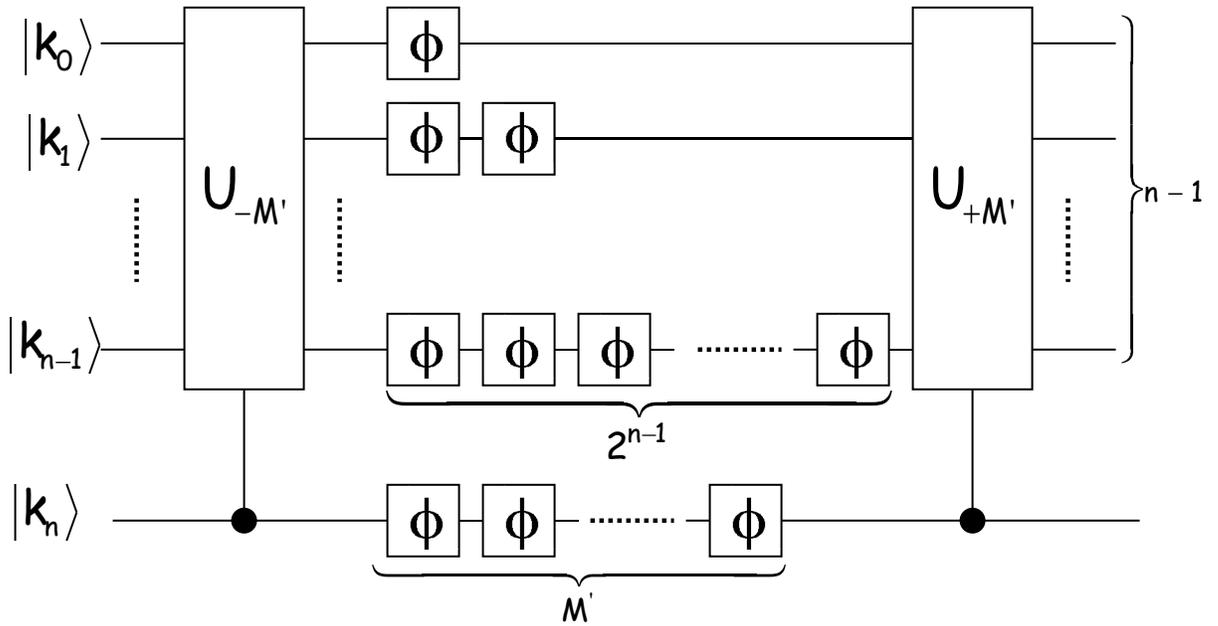} 
\caption{A quantum network for implementing the phase shift operator
$\ket{k} \mapsto e^{i \phi k} \ket{k}$ for $k \in \{0,1,\dots,
M\}$  represented in binary (that is $k = k_0 + 2k_1 + \cdots + k_n
2^n$) and $M < 2^{n+1}$. We let $M^{\prime} = 2^n - 1 -M$, and we
let $U_x$ denote the operator that maps $\ket{j} \mapsto 
\ket{j + x \bmod{2^n}}$.}
\label{f:exponentiate}
\end{center}
\end{figure*}

The network illustrated in Fig.~\ref{f:exponentiate} will achieve the
phase shift operator $\ket{k} \mapsto e^{i \phi k} \ket{k}$ where $k$
is represented in binary as the string of bits $k_0 k_1 \dots k_{n-1}
k_n$, where $k = k_0 + 2 k_1 + \cdots + 2^{n-1} k_{n-1}+2^n k_n$.

\section{Optimization of Fidelity Error}
\label{ap:maxfid}

In this appendix we derive the minimum obtainable value of the fidelity
error, given by
\begin{align}
\bar{C}_F & =  \frac{1}{2\pi} \sum_{y=0}^{M-1}
{{\int_{\phi=0}^{2\pi}{
\Prob(y|\phi)\cdot C_F(\phi,\tilde{\phi}_y)  d\phi}}}\\
& = \frac{1}{2}- \frac{1}{2}{\mathrm{Re}} \left(
{\sum_{j=0}^{N-1}{\alpha_j \alpha_{j+1}^*}} \right).
\label{eq:max}
\end{align}
We start by decomposing the complex coefficients $\alpha_j$ into the
two real parameters $p,q$ according to $\alpha_j = p_j +
q_j\sqrt{-1}$ for every $0\leq j\leq N$. This means that we have to
minimize the expression
\begin{equation} \label{eq:maxfn}
1-2\bar{C}_F  =  \sum_{j=0}^{N-1}{p_j p_{j+1}} +
\sum_{j=0}^{N-1}{q_j q_{j+1}}~,
\end{equation}
under the normalisation condition for the reals $p_j$ and $q_j$ that
\begin{equation} \label{eq:norm}
\sum_{j=0}^N{p_j^2} + \sum_{j=0}^N{q_j^2}   =  1~.
\end{equation}
First we will optimize the $p$ coefficients for the general case
$0\leq \sum{p_j^2} = \mu \leq 1$. We do this by the introduction of
a \emph{Lagrange multiplier\/} $\lambda$ and two additional values
$p_{-1} = p_{N+1} = 0$. The method of Langrange multipliers requires
that the partial derivatives $\partial/\partial p_t$ of the function
$\sum{p_j p_{j+1}} + \lambda \sum{p_j^2}$ have to be zero. This
leads to the set of equations for all $0\leq t \leq N$:
\begin{equation}
\frac{\partial}{\partial p_t} \left({\sum_{j=-1}^{N+1}{p_j p_{j+1}}
+ \lambda \sum_{j=-1}^{N+1}{p_j^2}}\right) = 0~,
\end{equation}
and hence the recurrence relation:
\begin{equation}
p_{t+1} =  -2\lambda p_t - p_{t-1}~.
\end{equation}
This relation can be solved with the help of the Chebyshev
polynomials of the second kind (see \cite{TABLE} Chap.~8.94,
p.~1032), using the identification $U_t(-\lambda) = p_t$. When
taking into account the restrictions $p_{-1}=p_{N+1}=0$ and
$\sum{p_j^2}=\mu$, the possible solutions are of the form
\begin{equation}
p_t  =  \pm \sqrt{\frac{2\mu}{N+2}}
\sin\left({\frac{(t+1)k\pi}{N+2}}\right)
\end{equation}
for $k \in \{1,2,\dots,N+1\}$. For each of the possible values of
$k$, the summation that we want to maximize equals (using
\cite{TABLE} Chap.~1.361, p.~32):
\begin{equation}
\sum_{t=0}^{N-1}{p_t p_{t+1}}  =  \mu \cos\left({ \frac{k\pi}{N+2}
}\right)~.
\end{equation}
The absolute maximum hence occurs when $k=1$ and equals
$\mu\cos(\pi/N+2)$. By the normalization condition of Equation
(\ref{eq:norm}) we have for the $q$ values that $\sum q_j^2 =
1-\mu$. Hence, the absolute maximum of the $\sum q_jq_{j+1}$
summation equals $(1-\mu)\cos(\pi/N+2)$. The overall result is
therefore the following.

The maximum $1-2\bar{C}_F$ as in Equation (\ref{eq:maxfn}) is
obtained for the $\alpha$ values
\begin{equation}
\alpha_j = 
e^{\I\psi}\sqrt{\smfrac{2}{N+2}}\sin\left({\smfrac{(j+1)\pi}{N+2}}\right)
\end{equation}
for any (non relevant) general phase factor $e^{\I\psi}$. The
investigated minimum then equals:
\begin{equation}
\bar{C}_F  = 
\sin^2\left({\smfrac{\pi}{2N+4}}\right)~,
\end{equation}
which gives $\bar{C}_F = O(1/N^2)$ as $N$ goes to infinity.

\section{Why a uniform prior?} \label{app_uniform_prior}

In this paper, we assign a uniform a priori probability for the value
of $\phi$. This is important for the results in this section.  There
are very natural situations in which one should assume a uniform prior
distribution, such as the case when there is no prior knowledge of the
value of $\phi$.  Another scenario (which is typical in computer
science) where one is naturally led to consider the uniform prior is
in an \emph{adversarial} scenario, where one party, Alice, is doing
the phase estimation, and the adversary, Bob, picks the phase. Alice
first describes the estimation procedure, and then Bob picks the phase
$\phi$ according to any probability distribution he wishes, with the
intention of maximizing the expected cost for Alice.

Alice must optimize her procedure to work well for any probability
distribution for $\phi$, since Bob will naturally pick the worst
case distribution for $\phi$. Alice can foil Bob's intentions by
``uniformizing'' the distribution of the $\phi$. That is, she can
guarantee that regardless of the distribution for $\phi$, her
procedure will work as well as it does for the uniform distribution
for $\phi$. That is, she can guarantee that the expected cost for
any distribution will equal the expected cost for the uniform prior
distribution.

Alice first takes any procedure $A$ that is optimal given a uniform
prior distribution for $\phi$.  She then augments this procedure in
the following way.  She pick a random phase $\phi_r$ with uniform
probability  over $[0,2\pi)$. She then runs the optimal procedure
for estimating $\phi$ but replaces each instance of $u_\phi$ with
$u_{\phi + \phi_r}$. This is easily done by adding a $u_{\phi_r}$
gate after every instance of a $u_\phi$ gate.  The procedure $A$
will output an estimate $\widetilde{{u_{\phi+\phi_r}}}$ for
$u_{\phi+\phi_r}$. Alice outputs $\widetilde{\widetilde{{u_{\phi}}}}
= \widetilde{{u_{\phi+\phi_r}}} - \phi_r $ as her estimate of
$\phi$.

This uniformization guarantees that regardless of the distribution
for $\phi$ that Bob chooses, Alice's new procedure performs with the
same expected cost as the optimal procedure would with a uniform
prior distribution.

However, if Alice is restricted to using finite means, she may not
be able to actually sample a uniform prior on the continuous set
$[0,2 \pi)$.  However, she can sample an arbitrarily fine discrete
subset of these points, which should be enough for dealing with
non-pathological cost functions $C$. Note that regardless of how
fine a mesh of points Alice samples from to ultimately produce
estimates $\phi^*$, the values of $\phi^*$ will come from a finite
set. Thus it is possible for an adversary to restrict to choices of
$\phi$ so that although the sets of numbers $\{ \phi - \phi^* \}$
are arbitrarily close for different $\phi$, they are still disjoint
for different $\phi$. This allows for the construction of
pathological cost functions for which even the slightest round-off
in the outputs $\phi^*$ can drastically change the expected cost.
Thus we need to add further practical assumptions for $C$ in the
case that we don't wish to assume an a priori uniform distribution
for $\phi$, but wish to justify a uniform prior distribution using
an adversarial scenario. Essentially, it suffices that most of the
time the function does not change very fast. More precisely, it
suffices for example that $|C(\phi)|<B < \infty$, and that for any
$\delta>0$ there exists an integer $N_{\delta}$ such that cost
function doesn't vary by more than $\delta$ over intervals of width
less than $1/N_{\delta}$, except for a set of points of measure at
most $\delta/B$. For example, any uniformly continuous cost function
satisfies this. Any bounded function that increases monotonically as
$\phi$ tends away from $0$ also satisfies this.

Knowing the details of the specific procedure she wishes to
uniformize, and knowing the cost function $C$, which we assume is
reasonably well-behaved as discussed above, and given any $\epsilon
> 0$, she can pick a large integer $D_{\epsilon}$ so that uniformly sampling
the phases $\{ 2\pi x/D_{\epsilon}| x=0,1,\dots, D_{\epsilon}-1\}$
will yield a procedure with expected cost within $\epsilon$ of the
optimal expected cost in the case of a perfectly uniform prior
distribution.

The above discussion is only meant to justify that a uniform prior
distribution is a meaningful assumption to make. In this paper, we
will simply assume that the prior distribution for the $\phi$ is
perfectly uniform in the interval $[0, 2\pi)$.

WvD is supported by the Disruptive Technology Office (DTO)
under Army Research Office (ARO) contract number W911NF-04-R-0009.
MM is supported by DTO-ARO, NSERC, CFI, ORDCF, CIAR, CRC, ORF,
and Ontario-MRI. This work was also supported in part by MIUR through
PRIN 2005 and by the EC through the project SECOQC.

\begin{thebibliography}{99}
\bibitem{nota}
This class of mixed states $\theta$ is characterised by the
restriction that the elements of $\theta$ have fixed phase along the
diagonals, i.e.\ $\theta_{hk}=|\theta_{hk}|e^{i\phi_{|h-k|}}$. In
this case the optimal $\chi$ is given by
$\chi_{hk}=e^{-i\phi_{|h-k|}}$.
\end{thebibliography}
\end{document}